\documentclass[aps,pra,twocolumn,amsmath,amssymb,amsfonts,floatfix]{revtex4}
\usepackage{bm}
\usepackage{bbm}

\def\ket#1{|#1\rangle }
\def\bra#1{\langle #1 |}
\def\braket#1{\langle #1 \rangle}
\def\n{\nonumber \\ }

\begin{document}

\title{Manifestation of topological behaviors in interacting Weyl systems: \\ one-body verse two-body correlations}

\author{Min-Fong Yang}
\affiliation{Department of Applied Physics, Tunghai University, Taichung 40704, Taiwan}

\date{\today}

\begin{abstract}
Understanding correlation effects in topological phases of matter is at the forefront of current research in condensed matter physics. Here we try to clarify some subtleties in studying topological behaviors of interacting Weyl semimetals.
It is well-known that there exist two topological invariants defined to identify their topological character. One is the many-body Chern number, which can be directly linked to the Hall conductivity and thus to the two-particle correlations. The other is the topological index constructed from the single-particle Green's functions. Because the information of Green's functions is easier to be achieved than the many-body wavefunctions, usually only the latter is employed in the literature.
However, the approach based on the single-particle Green's function can break down in the strongly correlated phase. For illustration, an exactly solvable two-orbital model with momentum-local two-body interactions is discussed, in which both topological invariants can be calculated analytically. We find that the topological index calculated from the Green's function formalism can be nonzero even for a non-topological strongly correlated phase with vanishing many-body Chern number. In addition, we stress that the physical surface states implied by nonzero many-body Chern numbers should be the edge modes of particle-hole collective excitations, rather than those of quasiparticle nature derived from the Green's function formalism.
Our observations thus demonstrate the limitation of the validity of Green's function formalism in the investigations of interacting topological materials.
\end{abstract}

\maketitle

\section{Introduction}

Weyl semimetals (WSMs) have received enormous attention
in contemporary research, as they broaden the classification of topological phases of matter beyond insulators~\cite{Hasan_etal2017,Yan-Felser2017,Burkov2018,Armitage_etal2018}. The band structures of these gapless three-dimensional materials display isolated band-touching points, called Weyl nodes. Such nodal points behave as (anti-)monopoles of the Berry curvature in the momentum space and carry
nonzero Chern numbers, which is proportional to the quantized Berry flux around the nodes.
Due to the bulk-boundary correspondence, the associated Chern number guarantees the existence of chiral surface states on the surfaces of a WSM~\cite{Wan_etal2011}. These surface states form open Fermi arcs terminating at the projections of two Weyl nodes with opposite Chern numbers onto the surface Brillouin zone. Recently, bulk Weyl nodes and surface Fermi arcs in TaAs and NbAs have both been observed in angle-resolved photoemission spectroscopy (ARPES) measurements~\cite{Xu_etal2015_1,Xu_etal2015_2,Lv_etal2015,Yang_etal2015}. Because of the nontrivial topological nature of Weyl nodes, WSMs are predicted to give many fascinating electromagnetic responses, such as chiral anomaly and other anomaly-induced transport phenomena~\cite{Hosur-Qi2013,Burkov_2015}.

Much of the novel physics in WSMs can be understood already at the level of a noninteracting description. Nonetheless, it is interesting to further investigate whether the aforementioned topological characters are robust against electronic correlations.
To answer this question, we need to generalize the concept of noninteracting Berry curvature to the correlated systems. Following the proposals in the context of interacting topological insulators~\cite{Wang-Qi-Zhang2010,Wang-Qi-Zhang2012,Wang-Zhang2012,Wang-Yan2013}, the Berry curvature constructed by the eigenstates of the zero-frequency single-particle Green's function is applied as well to the case of interacting WSMs~\cite{Witczak-Krempa_etal2014}.
Within cluster perturbation theory for the Green's function, it was shown that the essential features of WSMs persist (up to band renormalization) even in an interacting setting. This conclusion is further supported by analyzing the structure and symmetry of the diagrammatic expansion for the Green's function~\cite{Carlstrom-Bergholtz2018-1,Carlstrom-Bergholtz2018-2}. However, perturbative stability, even to infinite order, does not rule out the possibility of gap opening resulting from nonperturbative effects.

When either the spin density wave~\cite{Laubach_etal2016} or the Fulde-Ferrell-Larkin-Ovchinnikov superconducting~\cite{Wang_etal19} instabilities is allowed, based on some kinds of mean-field approximations, it was shown that a quantum anomalous Hall state can be stabilized, where the bulk Weyl nodes are gapped out but the Fermi arcs remain. This insulating phase is achieved by pair annihilation of Weyl nodes due to the backfolding of the Brillouin zone caused by ordering. Therefore, some special conditions either in the ordering vector or in the locations of a pair of Weyl nodes are required for the stability of this novel phase. Notably, these results show that, while a close relationship between the existence of the gapless Weyl nodes and the surface Fermi arc appears in the noninteracting regime, such a relationship needs not hold in the correlated systems.

Recently, the WSMs with two-body interactions that are local in momentum space, and then of infinite range in real space, are discussed~\cite{Morimoto-Nagaosa2016,Meng-Budich2019}. The advantage of such models is that different momentum states are decoupled and they thus can be solved exactly to uncover possible novel physics in the strongly correlated phases.
It is shown that the gap due to electron correlation can be opened at each Weyl node without the pair annihilation, in contrast to the mechanism proposed in Refs.~\cite{Laubach_etal2016,Wang_etal19}. Interestingly, the monopole charge of the Weyl node is unchanged due to the role of poles in Green's function being replaced by its zeros. The topological properties are thus kept unchanged: the Fermi arcs on the surfaces still persist and the Hall conductivity remains the same~\cite{Morimoto-Nagaosa2016}.
Moreover, when the momentum-local interactions is restricted to be nonvanishing around a single node, such that only one node is gapped out,
a WSM with unpaired gapless Weyl nodes can be realized~\cite{Meng-Budich2019}. This is forbidden by the fermion doubling theorem in noninteracting systems~\cite{nielsen_ninomiya-1,nielsen_ninomiya-2}. Remarkably, the chiral zeroth Landau level structure entailing the chiral anomaly survives even around the gapped node. However, there exists no zero-energy Fermi-arc surface state around the Weyl node once it is gapped by interactions. We note that the last conclusion is different from that in Ref.~\cite{Morimoto-Nagaosa2016}.

In this paper, some subtleties in studying interacting WSMs are discussed. We note that there exist two topological invariants available to identify the topological characters of interacting WSMs. One is the many-body Chern number $C$ [see Eq.~\eqref{topo1}], which measures the topology of the ground-state wavefunction. Because the Green's function is easier to be calculated than the many-body wavefunction, another topological index $\tilde{C}$ [see Eq.~\eqref{topo2_inG}] defined by the single-particle Green's function is more commonly employed~\cite{Witczak-Krempa_etal2014,Morimoto-Nagaosa2016,%
Meng-Budich2019,Acheche_etal2019,Zhang_Zubkov19,Xu_etal19}. We stress that these two topological invariants involve different physical processes and their values need not be equal in general. The many-body Chern number can be directly related to the Hall conductivity and thus contains the information of two-body correlations. By contrast, the topological index reflects the topological properties within the one-particle or one-hole subspace only.
As a consequence, the surface modes corresponding to these two different topological invariants must have distinct characters in physics. The physical surface states implied by nonzero many-body Chern numbers are the edge modes of particle-hole collective excitations, while those derived from the Green's function formalism are of quasiparticle nature. Due to the lack of clear physical meaning for the topological index $\tilde{C}$, conclusions drawn from this quantity could be misleading. This observation explains the inconsistency in the Fermi-arc surface states in Refs.~\cite{Morimoto-Nagaosa2016,Meng-Budich2019}.

To provide a concrete example demonstrating the failure of the predictions from the topological index $\tilde{C}$, an exactly solvable two-orbital model with momentum-local two-body interactions is studied. In the strong-correlation limit, the ground state is formed by spin-singlet s-wave pairs among two orbital states [see Eq.~\eqref{ground}] and the single-particle excitation energies are gapful. Besides, the many-body Chern number $C$ vanishes because the physical processes in creating the associated particle-hole excitations are forbidden. The system thus behaves as a non-topological Mott insulator with a full gap. However, due to the contributions from the zero, rather than the physical pole, of the single-particle Green's function, the topological index $\tilde{C}$ remains nonzero. This clearly shows that the Green's function approach can give wrong predictions to the topological behaviors of the strongly correlated systems.

This paper is organized as follows. In Sec. II, two kinds of topological invariants are introduced and the difference in their physical content is explained. In Sec. III, an exactly solvable model of interacting WSMs with a momentum-local pair-hopping term is discussed to demonstrate the inequivalence of these two topological invariants. We summarize our work in Sec. IV. To be self-contained, general Green's functions in the Matsubara formalism and the solvable Morimoto-Nagaosa model~\cite{Morimoto-Nagaosa2016} are briefly reviewed in Appendices A and B, respectively.

\section{Two kinds of topological invariants}

For simplicity, we consider the WSMs hosting a single pair of Weyl nodes separated by a distance along the $k_z$ direction in the absence of interaction. In such a situation, the region between nodes can be viewed as a stack of two-dimensional Chern insulators.
As pointed out in the context of interacting topological insulators, for each two-dimensional momentum sector (say, for a fixed $k_z$),  we can define two kinds of topological invariants~\cite{Wang-Qi-Zhang2010}. One is the Chern number $C(k_z)$ [see Eq.~\eqref{topo1}], which is directly related to the Hall conductance. The other is the topological index $\tilde{C}(k_z)$ [see Eq.~\eqref{topo2_inG}] measuring the homotopy of the single-particle Green's function in combined frequency-momentum space. While similar in their expressions, the physics implied by these two topological invariants is different~\cite{Budich-Trauzettel2013,Hohenadler-Assaad2013,Meng_etal2014,Rachel18}, as explained below.

\subsection{Chern number $C$}
\label{Chern_num}

As mentioned above, our system can be considered as a stack of two-dimensional models. For each two-dimensional system of fixed $k_z$, the many-body Chern number can be obtained through an integration over the twist angles $\theta_\mu\in[0,2\pi)$ ($\mu=x$, $y$) of general boundary conditions~\cite{Niu_etal1985},
\begin{align}\label{topo1}
C(k_z)= 2\pi i \int\int^{2\pi}_{0} \frac{d\theta_x d\theta_y}{(2\pi)^2}
\left[ \left\langle\frac{\partial\Psi_0}{\partial\theta_x}\Bigg|\frac{\partial\Psi_0}{\partial\theta_y}\right\rangle - (x\leftrightarrow y) \right] \; .
\end{align}
Here $\ket{\Psi_0}$ denotes the exact ground state of the many-body Hamiltonian $H$ with the energy eigenvalue $E_0$ and is assumed to be non-degenerate. The integrand represents the many-body counterpart of the Berry curvature. We note that this Berry curvature is constructed by the many-body ground-state wavefunction, which may not be expressed as a product of single-particle wavefunctions in the case of interacting systems.

In the thermodynamic limit, twisted boundary conditions do not affect bulk properties and the integration in calculating the many-body Chern number is thus unnecessary~\cite{Niu_etal1985,Hastings-Michalakis2015,Kudo_etal2019}. That is, it suffices to compute the Chern number for a single set of the twist angles, say, for the periodic boundary condition with $\theta_x=0$ and $\theta_y=0$. In that case and for systems with translational invariance, one can further replace the derivatives with respect to the twist angles by momentum derivatives~\cite{Niu_etal1985}, i.e., $\partial/\partial\theta_\mu\rightarrow(1/L)\partial/\partial k_\mu$, where $L$ denotes the linear size of systems.

When the excitation gap is nonvanishing, the many-body Chern number can be rewritten as~\cite{Niu_etal1985}
\begin{align}\label{kubo}
C(k_z)=\frac{2\pi i}{L^2} \sum_{n>0} \frac{\langle\Psi_0|v_x|\Psi_n\rangle \langle\Psi_n|v_y|\Psi_0\rangle - (x\leftrightarrow y)}{(E_n-E_0)^2/\hbar^2}
\end{align}
with $v_\mu=(1/\hbar)\,\partial H/\partial k_\mu$ ($\mu=x$, $y$) being the velocity operator and $\ket{\Psi_n}$ the $n$-th excited state with the energy $E_n$. This expression reveals the relation to the Kubo formula of the Hall conductivity for two-dimensional systems (for example, see App.~\ref{Green}). Therefore, for the present three-dimensional interacting WSMs, the anomalous Hall conductivity $\sigma_{xy}$ becomes the sum over the values for different $k_z$ sectors,
\begin{equation}
\sigma_{xy}= \int^\pi_{-\pi} \frac{dk_z}{2\pi} \frac{e^2}{h} C(k_z) \; .
\end{equation}

Due to the close relation between the Chern number and the Hall conductivity, two remarks on the Chern number and the associated Fermi arc states are discussed.
First, the Hall conductivity and hence the Chern number arise from the particle-hole excitations and reflect \emph{two-particle} correlations~\cite{Mahan}. They thus may not always be decomposed into single-particle processes. This observation can be understood because the Hall conductivity stems from the current-current correlation function and the current (or velocity) operator contains a product of one creation and one annihilation operator.
Second, the surface Fermi arc states implied by nonzero Hall conductivity (and Chern number) correspond to \emph{bosonic charge-neutral} particle-hole excitations. That is, they are collective edge excitations within the $\Delta N=0$ subspace of the Hilbert space, instead of charged single-particle edge modes with $\Delta N=\pm 1$, where $\Delta N$ denotes the change in total particle number. This conclusion results from the theory of quantum Hall edge states~\cite{Wen1991,Renn1995,Yoshioka}. According to that thoery, chiral edge modes for each two-dimensional Chern insulator of a fixed $k_z$ should be capillary waves, whose dynamics is described by a one-dimensional chiral boson theory. In short, the physics implied by nonzero many-body Chern number is not of single-particle nature as those inferred from the Green's function formalism (see the next subsection).

From the above observations, one can easily explain the novel result found in the previous solvable model of interacting WSMs~\cite{Morimoto-Nagaosa2016}. It was shown that the Hall conductivity (and thus many-body Chern number) preserves its noninteracting value even though both Weyl nodes are gapped out. This puzzling conclusion can be resolved by noticing that the physical processes relevant to the Hall conductivity are the particle-hole excitations, instead of those of quasiparticle nature. As elucidated in App.~\ref{MNmodel}, while all the single-particle/hole excitations have finite excitation gaps, the bosonic charge-neutral particle-hole excitations can keep their noninteracting values of excitation energies, which reduce to zero at Weyl nodes. Besides, the velocity operator and its matrix elements all take the same forms as those in the noninteracting case. Namely, within the particle-hole subspace, the effective theory for calculating the Hall conductivity is identical to an interaction-free model with gapless nodes. This explains how the Hall conductivity can survive even in the presence of interactions.

\subsection{Topological index $\tilde{C}$}
\label{top_index}

Because the evaluation of the many-body Chern number for the correlated systems can be quite involved, the authors in Ref.~\cite{Wang-Qi-Zhang2010} proposed another topological index $\tilde{C}$ (or $N_2$ if one follows their notation) to replace the role played by the Chern number. For the plane of fixed $k_z$, it is defined by~\cite{note_top2}
\begin{align}\label{topo2_inG}
\tilde{C}(k_z)=\;&\frac{\epsilon^{\mu\nu\rho}}{6} \int_{-\infty}^{\infty} dk_0 \int \frac{dk_x dk_y}{(2\pi)^2} \nonumber \\
&\times\mathrm{tr} \Big[ G(\partial_\mu G^{-1}) G(\partial_\nu G^{-1})
   G(\partial_\rho G^{-1}) \Big] \; .
\end{align}
Here $G=G(\mathbf{k},i\omega)$ is the exact single-particle Green's function, $\partial_\mu\equiv\partial/\partial k_\mu$ with $\mu$, $\nu$, $\rho\in\{0,x,y\}$ and $k_0=i\omega$. $\epsilon_{\mu\nu\rho}$ is the totally antisymmetric tensor, and the operation $\mathrm{tr}$ indicates the trace over the Green's function matrix structure. A summation over $\mu$, $\nu$, and $\rho$ is implied. This quantity shows the homotopy class of the single-particle Green's function $G$ in combined frequency-momentum space. However, because involving an integration in the $i\omega$ direction, it is not easy to evaluate
the topological index $\tilde{C}$ through this expression.

Interestingly, the topological information encoded by $\tilde{C}(k_z)$ can be obtained as well by the exact single-particle Green's function at zero frequency~\cite{Wang-Zhang2012,Wang-Yan2013}. If one define an effective Bloch Hamiltonian,
\begin{equation}\label{topo_Hami}
\mathcal{H}_t \equiv -G^{-1}(\mathbf{k},i\omega=0) \; ,
\end{equation}
which is dubbed as the topological Hamiltonian~\cite{Wang-Yan2013}, the topological index $\tilde{C}(k_z)$ can then be rewritten as an integral of a Berry curvature constructed via the eigenstates of $\mathcal{H}_t$. That is, denoting the eigenstates of $\mathcal{H}_t$ by $\ket{\phi_n(\mathbf{k})}$ with the band structure $\{\xi_n(\mathbf{k})\}$, we have
\begin{align}\label{topo2}
\tilde{C}(k_z)=i&\int^\pi_{-\pi}\int^\pi_{-\pi} \frac{dk_x dk_y}{2\pi}
\sideset{}{'}\sum_{n} \nonumber \\
&\times \left[ \left\langle\frac{\partial\phi_n(\mathbf{k})}{\partial k_x}\Bigg|\frac{\partial\phi_n(\mathbf{k})}{\partial k_y}\right\rangle - (x\leftrightarrow y) \right] \; .
\end{align}
Here the primed summation goes over the states with nonpositive eigenvalues, i.e., $\xi_n(\mathbf{k})\leq0$, which correspond to effective occupied states. It is clear that these eigenstates can be loosely viewed as substitutes of the Bloch states of a noninteracting system.
Once the effective single-particle Hamiltonian $\mathcal{H}_t$ is obtained, the procedure in calculating $\tilde{C}(k_z)$ becomes the same as the noninteracting case. It is the reason why this topological index is usually employed in the studies on interacting WSMs.

While different in their expressions, as discussed in Ref.~\cite{Wang-Qi-Zhang2010} 
(see also Sec.~6.4 of Ref.~\cite{Budich-Trauzettel2013}),
the values of the topological index $\tilde{C}(k_z)$ and the Chern number $C(k_z)$ can be the same if the interacting phase can be adiabatically connected to a non-interacting counterpart without gap closing. Otherwise, their equality may break down for some strongly correlated phases. In that case, finite $\tilde{C}(k_z)$ does not guarantee the Chern number $C(k_z)$ to be nonzero, and then the conclusions drawn merely from the topological index $\tilde{C}(k_z)$ should be considered with caution.

Moreover, the bulk-boundary correspondence for the topological index $\tilde{C}(k_z)$ can fail in the presence of electron correlations~\cite{Gurarie2011,Essin-Gurarie2011}. By using the identity, $G(\partial_\mu G^{-1})=-(\partial_\mu G)G^{-1}$, one finds that the role played by the Green's function and its inverse can be interchanged, as seen from Eq.~\eqref{topo2_inG}. This implies that, besides the physical poles in the Green's function, zeros in $G$ (i.e., poles in $G^{-1}$) can also contribute to the topological index $\tilde{C}(k_z)$. However, as explained by the authors in Refs.~\cite{Gurarie2011,Essin-Gurarie2011}, such Green function zeros do not guarantee the existence of physical edge states, while they do give nonzero contribution to $\tilde{C}(k_z)$. Therefore, even though one can derive the edge modes by solving the effective single-particle Hamiltonian $\mathcal{H}_t$ under open boundaries, these modes may not correspond to physical states.

The last observation explains the disagreement in the Fermi arc states of the solvable two-band model of interacting WSMs~\cite{Morimoto-Nagaosa2016,Meng-Budich2019}. For this solvable model, it is shown that the topological index $\tilde{C}(k_z)$ preserves its noninteracting value and the topological Hamiltonian $\mathcal{H}_t$ has the same form as the noninteracting one (see also App.~\ref{MNmodel}). The nonzero $\tilde{C}(k_z)$ seems to imply the existence of Fermi arc states. In Ref.~\cite{Morimoto-Nagaosa2016} (see its supplementary information), such single-particle surface states are explicitly derived by solving $\mathcal{H}_t$ under open boundary conditions.
However, the eigenstates of the effective single-particle Hamiltonian $\mathcal{H}_t$ have no direct physical meaning, and the Fermi arc states derived from $\mathcal{H}_t$ thus may not be physical observable. Indeed, by directly diagonalizing the many-body Hamiltonian $H$, instead of $\mathcal{H}_t$, within the subspace of single-particle excitations and under open boundary conditions, no single-particle surface mode at zero energy is found when both Weyl nodes are gapped out~\cite{Meng-Budich2019}.

\section{exactly solvable two-orbital model of interacting WSMs}

In this section, an exactly solvable two-orbital model of WSMs with momentum-local inter-orbital interaction is explored, in which the many-body Chern number $C(k_z)$ and the topological index $\tilde{C}(k_z)$ can be evaluated analytically.
In contrast to the case of Morimoto-Nagaosa model~\cite{Morimoto-Nagaosa2016} (see also App.~\ref{MNmodel}), the equivalence of two topological invariants does break down in the present model.
This explicitly demonstrates that the knowledge of the topological index $\tilde{C}(k_z)$ can sometimes give completely wrong understanding on the topological properties of generic interacting systems.

\subsection{Model Hamiltonian}

Our model Hamiltonian with momentum-local inter-orbital interaction $V$ ($V>0$) is given by
\begin{align}\label{2orb_model}
H &= \sum_{\mathbf{k},\alpha} \psi^\dagger_{\mathbf{k}\alpha} \left[\mathbf{h}(\mathbf{k})\cdot\bm{\sigma}\right]
\psi_{\mathbf{k}\alpha} \n
&\quad + \frac{V}{2} \sum_\mathbf{k} \left[ \left( \psi^\dagger_{\mathbf{k}1} \psi_{\mathbf{k}2} \right)^2 + \left( \psi^\dagger_{\mathbf{k}2} \psi_{\mathbf{k}1} \right)^2 \right] \; ,
\end{align}
where $\alpha=1$, 2 denotes the orbital index and $\psi^\dagger_{\mathbf{k}\alpha}=(c^\dagger_{\mathbf{k}\alpha\uparrow},\; c^\dagger_{\mathbf{k}\alpha\downarrow})$ the spinor of creation operators of electrons of (pseudo-)spin $\sigma=\;\uparrow$, $\downarrow$.
The first part describes a minimal two-band model of a Weyl semimetal with orbital degeneracy. Here only a single pair of Weyl nodes located at momenta $\pm\mathbf{k}_\mathrm{Weyl}=(0,\,0,\,\pm k_\mathrm{Weyl})$ is assumed for simplicity.
The second part represents the pair hopping from one orbital state to the other, but keeping $\mathbf{k}$ and $\sigma$ being unchanged [see also Eq.~\eqref{2orb_model_v2}]. Note that the interaction term is local in $\mathbf{k}$ space rather than in real space.

After performing a unitary transformation $U_\mathbf{k}$ for both orbital states to change to a new spinor $\phi^\dag_{\mathbf{k}\alpha}=\psi^\dag_{\mathbf{k}\alpha}\,U_\mathbf{k}
=(b^\dag_{\mathbf{k}\alpha+},\; b^\dag_{\mathbf{k}\alpha-})$ such that $U_\mathbf{k}^\dag\,[\mathbf{h}(\mathbf{k})\cdot\bm{\sigma}]\,U_\mathbf{k} = \sigma_z h(\mathbf{k})$
with $h(\mathbf{k})=|\mathbf{h}(\mathbf{k})|$, the Hamiltonian becomes
\begin{align}\label{2orb_model_v2}
H &= \sum_{\mathbf{k},\alpha} h(\mathbf{k})\,\left( b_{\mathbf{k}\alpha+}^\dag b_{\mathbf{k}\alpha+} - b_{\mathbf{k}\alpha-}^\dag b_{\mathbf{k}\alpha-} \right)  \nonumber \\
&\quad + V \sum_\mathbf{k} \left( b_{\mathbf{k}1+}^\dag b_{\mathbf{k}2+} b_{\mathbf{k}1-}^\dag b_{\mathbf{k}2-} +  b_{\mathbf{k}2+}^\dag b_{\mathbf{k}1+} b_{\mathbf{k}2-}^\dag b_{\mathbf{k}1-} \right)  \; .
\end{align}
Because different momentum states are decoupled, this interacting model can thus be solved exactly through diagonalization for each $\mathbf{k}$. Notice that the total particle number for each $\mathbf{k}$ with a given (pseudo-) spin state of the new basis, $n_{\mathbf{k}\sigma}=\sum_\alpha b_{\mathbf{k}\alpha\sigma}^\dag b_{\mathbf{k}\alpha\sigma}$, commutes with the Hamiltonian $H$. The energy eigenstates of $H$ will thus be the eigenstates of $n_{\mathbf{k}\sigma}$ as well.

In the strong correlation limit such that $V>W$, where $W$ denotes the noninteracting bandwidth such that $W/2=\max_\mathbf{k}(h(\mathbf{k}))$, the many-body ground state at zero temperature and at half-filling becomes
\begin{equation}\label{ground}
|\Psi_0\rangle = \prod_\mathbf{k} \frac{1}{\sqrt{2}} (b_{\mathbf{k}1+}^\dag b_{\mathbf{k}1-}^\dag - b_{\mathbf{k}2+}^\dag b_{\mathbf{k}2-}^\dag)\ket{0} \; .
\end{equation}
Here $\ket 0$ denotes the vacuum state. This ground state consists of spin-singlet s-wave pairs among two orbital states. The single-particle/hole excited states, $b_{\mathbf{k}\alpha\sigma}^\dag|\Psi_0\rangle$ and $b_{\mathbf{k}\alpha\bar{\sigma}}|\Psi_0\rangle$ for $\sigma=\pm$ and $\bar{\sigma}=-\sigma$, have excitation energies $\Delta E=V+\sigma h(\mathbf{k})$, which are nonzero for all $\mathbf{k}$. The system thus behaves as a Mott insulator with a full gap. Therefore, no gapless mode will appear in the spectral function and can be observed in ARPES measurements. We note that this ground state cannot be adiabatically connected to that in the noninteracting case of $V=0$ without gap closing. Due to the absence of such an adiabatic connection, as discussed in Sec.~\ref{top_index}, the values of two topological invariants, $C(k_z)$ and $\tilde{C}(k_z)$, need not be the same for the present strongly correlated state. As explicitly shown below, we do find that the topological index $\tilde{C}(k_z)$ is nonzero, while the many-body Chern number $C(k_z)$ vanishes completely.

\subsection{Single-particle Green's function and topological index $\tilde{C}$}
\label{tildeC_2orb_model}

By using the exact eigenstates and the corresponding energy spectrum, the thermal single-particle Green's function can be written down, which becomes diagonal in the (pseudo-)spin space for the new basis. According to the general expression of the Green's function at zero temperature (see Eq.~\eqref{eq:Green0} in App.~\ref{Green}) with $\hat{A}=b_{\mathbf{k}\alpha\sigma}$ and $\hat{B}=b_{\mathbf{k}\alpha\sigma}^\dag$, we have
\begin{align}
&G_{\alpha\sigma;\alpha\sigma}(\mathbf{k},i\omega) \nonumber \\
&=\frac{1}{2} \left( \frac{1}{i\omega-\sigma[V+h(\mathbf{k})]}
 + \frac{1}{i\omega+\sigma[V-h(\mathbf{k})]}  \right) \; .
\end{align}
Therefore, away from the Weyl nodes such that $h(\mathbf{k})\neq 0$, the Green's function in the original basis becomes
\begin{align}\label{G_2orb_model}
G(\mathbf{k},i\omega) =\frac{1}{2} \left( \frac{1}{i\omega-\mathbf{h}_+(\mathbf{k})\cdot\bm{\sigma}}
+ \frac{1}{i\omega+\mathbf{h}_-(\mathbf{k})\cdot\bm{\sigma}} \right) \; ,
\end{align}
where $\mathbf{h}_\pm(\mathbf{k}) =
\mathbf{n}(\mathbf{k})[V\pm h(\mathbf{k})]$
with $\mathbf{n}(\mathbf{k})=\mathbf{h}(\mathbf{k})/h(\mathbf{k})$.
On the other hand, at the Weyl nodes with $h(\mathbf{k}=\pm\mathbf{k}_\mathrm{Weyl})=0$, we have
\begin{align}\label{Green_zero}
G(\mathbf{k}=\pm\mathbf{k}_\mathrm{Weyl},i\omega) %
&=\frac{1}{2} \left( \frac{1}{i\omega-V} + \frac{1}{i\omega+V} \right) \mathbbm{1}\;\nonumber \\
&=\frac{i\omega}{(i\omega)^2 - V^2}\;\mathbbm{1} \; ,
\end{align}
where $\mathbbm{1}$ denotes the $2\times2$ identity matrix in the (pseudo-)spin space. We note that the zero-energy poles at $\mathbf{k}=\pm\mathbf{k}_\mathrm{Weyl}$ in the noninteracting Green's function are now replaced by the zero-energy zeros for the present strongly correlated state.

As discussed in Sec.~\ref{top_index}, with the single-particle Green's function in hand, the topological index $\tilde{C}(k_z)$ of our solvable model is ready to be evaluated. We find that, while the gapless Weyl nodes are gapped out in the presence of $V$, the topological index $\tilde{C}(k_z)$ remains nonzero and equals to its noninteracting value with reversed sign.
As explained in Refs.~\cite{Gurarie2011,Essin-Gurarie2011}, this result can be easily understood because the zero-energy zeros in Eq.~\eqref{Green_zero} now take the place of the original Weyl nodes and behave as sources or drains of the Berry flux for $\tilde{C}(k_z)$ in Eq.~\eqref{topo2_inG}.
Our result of the topological index $\tilde{C}(k_z)$ can be obtained by using Eq.~\eqref{topo2} as well. Employing the single-particle Green's function in Eq.~\eqref{G_2orb_model}, the so-called topological Hamiltonian defined in Eq.~\eqref{topo_Hami} becomes
\begin{equation}\label{our_topo_Hami}
\mathcal{H}_t \equiv \mathbf{h}_\mathrm{eff}(\mathbf{k})\cdot\bm{\sigma} \; ,
\end{equation}
where $\mathbf{h}_\mathrm{eff}(\mathbf{k})%
=-\mathbf{n}(\mathbf{k})[V^2-(h(\mathbf{k}))^2]/h(\mathbf{k})$. This effective single-particle Hamiltonian is the same as the noninteracting Hamiltonian up to a negative multiplication factor. The eigenstates $\ket{\phi_n(\mathbf{k})}$ of $\mathcal{H}_t$ are thus identical to those in the noninteracting case, but the signs of their eigenvalues are reversed. Hence the topological index $\tilde{C}(k_z)$ calculated from Eq.~\eqref{topo2} remains nonzero but changes its sign in comparison to the case of $V=0$.

\subsection{Chern number $C$}

From our results of finite single-particle excitation gap and nonzero topological index $\tilde{C}(k_z)$, one may conclude that the strongly correlated phase for $V>W$ is a gapped Mott insulator with non-trivial topology. However, as calculated below, the many-body Chern number $C(k_z)$ vanishes completely. This indicates that our insulating phase should be non-topological instead.

For convenience, we calculate $C(k_z)$ by using Eq.~\eqref{kubo}, which is closely related to the Kubo formula of the Hall conductivity. The velocity operator for our model in Eq.~\eqref{2orb_model} is given by
\begin{equation}\label{eq:v_op}
v_\mu = \sum_\mathbf{k,\alpha}\sum_{\sigma,\sigma'=\pm} %
\frac{1}{\hbar} \left( \frac{\partial\mathcal{H}(\mathbf{k})}{\partial k_\mu}\right)_{\sigma\sigma'} %
b_{\mathbf{k}\alpha\sigma}^\dagger b_{\mathbf{k}\alpha\sigma'}
\end{equation}
with
\begin{equation}
\left( \frac{\partial\mathcal{H}(\mathbf{k})}{\partial k_\mu} \right)_{\sigma\sigma'} %
\equiv \left\langle\mathbf{k}\sigma\left| \frac{\partial\mathcal{H}(\mathbf{k})}{\partial k_\mu} \right|\mathbf{k}\sigma'\right\rangle \; ,
\end{equation}
where the single-particle states $\ket{\mathbf{k}\sigma}$ are the eigenstates of the single-particle Hamiltonian matrix $\mathcal{H}(\mathbf{k})=\mathbf{h}(\mathbf{k})\cdot\bm{\sigma}$. Because the energy eigenstate $\ket{\Psi_n}$ is also an eigenstate of the total particle number for a given $\mathbf{k}$ and $\sigma$, we have
$\bra{\Psi_n}\,\sum_\alpha b_{\mathbf{k}\alpha\sigma}^\dag b_{\mathbf{k}\alpha\sigma}\ket{\Psi_0} \propto \langle\Psi_n\ket{\Psi_0}=0$ for $\ket{\Psi_n}$ being the excited states. The possible contributions in the numerators of Eq.~\eqref{kubo} thus come from the terms involving the matrix elements with opposite (pseudo-)spin indices, i.e.,  $|\bra{\Psi_n}\,b_{\mathbf{k}\alpha+}^\dag\,b_{\mathbf{k}\alpha-}\ket{\Psi_0}|^2$ and
$|\bra{\Psi_n}\,b_{\mathbf{k}\alpha-}^\dag\,b_{\mathbf{k}\alpha+}\ket{\Psi_0}|^2$.
However, these particle-hole excitations are forbidden because both $b_{\mathbf{k}\alpha+}^\dag\,b_{\mathbf{k}\alpha-}\ket{\Psi_0}$ and
$b_{\mathbf{k}\alpha-}^\dag\,b_{\mathbf{k}\alpha+}\ket{\Psi_0}$ vanish for the ground state $\ket{\Psi_0}$ in Eq.~\eqref{ground}. That is, such bosonic charge-neutral particle-hole excitations in our strongly correlated phase are not allowed at all. Hence the strongly correlated phase of the present interacting WSMs has zero many-body Chern number and then carries no anomalous Hall effect.

\subsection{Fermi arc states}

Since the two topological invariants are shown to exhibit distinct behaviors in our strongly correlated phase, their implications on the corresponding surface states should be different. We now discuss the consequences of both topological invariants on the Fermi arc states.

Due to the appearance of zero-energy zeros in the single-particle Green's function, the topological index $\tilde{C}(k_z)$ remains nonzero in the $k_z$ region between the gapped nodes. According to the usual bulk-boundary correspondence, this seems to imply the existence of the single-particle edge states for those $k_z$ sectors and thereof the formation of the Fermi arcs on the surface Brillouin zone. In previous investigations on interacting WSMs, the Fermi arc states are sometimes calculated by considering the topological Hamiltonian $\mathcal{H}_t$ under some kinds of boundary conditions~\cite{Morimoto-Nagaosa2016,Xu_etal19}. As shown in Eq.~\eqref{our_topo_Hami}, $\mathcal{H}_t$ for the present model has the same form as the noninteracting Hamiltonian. This supports the survival of the noninteracting Fermi arc states in the strong correlation limit, even though both Weyl nodes are gapped out.

However, we stress that the eigenstates of the effective single-particle Hamiltonian $\mathcal{H}_t$ have no direct physical meaning. The single-particle Fermi arc states derived from $\mathcal{H}_t$ thus may not be physical observable. Actually, as pointed out in Refs.~\cite{Gurarie2011,Essin-Gurarie2011} in the context of topological insulators, Green function zeros are in general not related to the existence of physical edge states. The bulk-boundary correspondence can thus fail in the presence of electron correlations. The solvable two-band model mentioned in Sec.~\ref{top_index} provides a concrete example for the failure in predicting Fermi arc states by using the topological index $\tilde{C}(k_z)$ and the topological Hamiltonian $\mathcal{H}_t$. Therefore, the conclusions drawn from $\tilde{C}(k_z)$ and $\mathcal{H}_t$ should be considered with caution.

On the other hand, the many-body Chern number $C(k_z)$ always leads to physical consequences. It is thereof the correct indicator for the topological properties of interacting WSMs. Because of its relation to the Hall conductivity, according to the theory of quantum Hall edge states~\cite{Wen1991,Renn1995,Yoshioka}, the surface states inferred by nonzero Chern number should correspond to bosonic charge-neutral particle-hole excitations, instead of charged single-particle edge modes implied from the Green's function formalism. For our strongly correlated phase, as discussed in the last subsection, the Chern number is zero and the related collective particle-hole edge excitations are completely forbidden. We thus conclude that there exists no topological low-energy bosonic boundary mode in our strongly correlated phase.

\section{Conclusion}

In conclusion, to reexamine the equivalence of two topological invariants in Eqs.~\eqref{topo1} and~\eqref{topo2_inG}, an exactly solvable model of interacting two-orbital WSMs with a momentum-local pair-hopping term is discussed. Just like the noninteracting case, the topological index $\tilde{C}(k_z)$ is found to be nonzero between two Weyl nodes even though these nodes are gapped out in the strongly correlated phase. However, the many-body Chern number $C(k_z)$ vanishes for all values of $k_z$.  The disagreement between $\tilde{C}(k_z)$ and $C(k_z)$ may not be surprising since these two topological invariants encode difference types of correlations. Furthermore, the breakdown of their equivalence does not conflict with the common wisdom, because there exists no noninteracting correspondence for the ground state in the present strongly correlated phase.

We note that the topological index $\tilde{C}(k_z)$ based on the single-particle Green's function formalism can sometime give misleading conclusions, no matter whether its value is equal to the Chern number $C(k_z)$ or not. Due to the direct relevance to real processes, only the Chern number $C(k_z)$ represents the true physics.
First, in the present case, the topological index $\tilde{C}(k_z)$ can be nonzero even for a non-topological Mott insulating phase with a full gap and vanishing Chern number $C(k_z)$.
Second, for the solvable Morimoto-Nagaosa model~\cite{Morimoto-Nagaosa2016}, where both topological invariants are equal, the possible zero-energy single-particle surface states suggested by nonzero $\tilde{C}(k_z)$ do not show up in the single-particle spectrum obtained by exact diagonalization of the many-body Hamiltonian under open boundary conditions~\cite{Meng-Budich2019}.
Furthermore, similar observation on the inconsistency of the two topological invariants has been found in the context of interacting topological insulators~\cite{You_etal2014,He_etal2016}. For two specific two-dimensional bilayer models, it was shown that the topological invariant calculated from the Green's function formalism (i.e., the topological index in the present work) is finite, but no zero-energy edge state exists in the featureless Mott insulator phase~\cite{He_etal2016}. All these examples indicate that it could be dangerous to draw conclusions merely from the results based on the single-particle Green's function formalism.

Because most of the above discussions do not depend on the details of the non-interacting Hamiltonian, our conclusions can thus be applied to more general cases, such as multi-WSMs with a single pair or more pairs of Weyl nodes. Furthermore, our observations bring cautions as well to the investigations on other interacting quantum anomalous Hall systems, when only the topological invariant based on the single-particle Green's function formalism is calculated.

\begin{acknowledgments}
The author is grateful to M.-C. Chang for interesting and fruitful discussions. This work was supported from the Ministry of Science and Technology of Taiwan under Grant No. MOST 107-2112-M-029-004 and MOST 107-2112-M-029-006.
\end{acknowledgments}

\appendix
\section{Green's function in the Matsubara formalism}
\label{Green}

The imaginary-time Green's function at finite temperature for operators $\hat{A}$ and $\hat{B}$ is defined as
\begin{eqnarray}
\label{eq:MatsubaraGreenDef}
G_{\hat{A}\hat{B}}(\tau) = -\Big\langle T_{\tau}[\hat{A}(\tau)\hat{B}(0)] \Big\rangle \; ,
\end{eqnarray}
where the time-ordering symbol $T_{\tau}$ in imaginary time has been introduced and $\hat{A}(\tau)\equiv e^{\tau\hat{H}}\hat{A}e^{-\tau\hat{H}}$.

Using the Lehmann representation via expanding the expectation value by all the exact eigenstates $\{ |\Psi_n\rangle \}$ and the corresponding eigenvalues $\{ E_n \}$ of the many-body Hamiltonian $\hat{H}$, the Fourier transform of $G_{\hat{A}\hat{B}}(\tau)$ can be written as~\cite{Bruus-Flensberg}
\begin{align}
\label{eq:Transform}
G_{\hat{A}\hat{B}}(i\omega_\nu)&=\int_{0}^{\beta} G_{\hat{A}\hat{B}}(\tau) e^{i\omega_\nu\tau}d\tau \nonumber \\
&= \sum_{m,n} \frac{\langle\Psi_n|\hat{A}|\Psi_m\rangle %
\langle\Psi_m|\hat{B}|\Psi_n\rangle}{i\omega_\nu-(E_m-E_n)}
\frac{e^{-\beta E_n} \pm e^{-\beta E_m}}{Z} \; .
\end{align}
Here $\beta=1/k_B T$ is the inverse temperature and $Z=\mathrm{Tr}(e^{-\beta\hat{H}})$ is the partition function. The upper (lower) sign is for fermionic (bosonic) operators, which originates from the term $e^{i\omega_\nu\beta}=-1$ ($e^{i\omega_\nu\beta}=+1$) for fermionic (bosonic) Matsubara frequencies $\omega_\nu=(2\nu+1)\pi/\beta$ ($\omega_\nu=2\nu\pi/\beta$) with $\nu\in\mathbb{Z}$.

In the zero-temperature limit ($\beta\rightarrow\infty$), the above Green’s function reduces to
\begin{align}\label{eq:Green0}
G_{\hat{A}\hat{B}}(i\omega_\nu) &= \sum_{n\neq0} \left( \frac{\bra{\Psi_0}\hat{A}\ket{\Psi_n}\bra{\Psi_n}\hat{B}\ket{\Psi_0}}%
{i\omega - (E_n-E_0)} \right. \nonumber \\
& \left. \quad \pm
\frac{\bra{\Psi_0}\hat{B}\ket{\Psi_n}\bra{\Psi_n}\hat{A}\ket{\Psi_0}}%
{i\omega + (E_n-E_0)} \right) \, .
\end{align}

An application of the above formula is to derive the Hall conductivity $\sigma_H$ from the Kubo formula. The current-current correlation function $\Pi_{xy}(\mathbf{q}, \tau)$ in the limit of the momentum transfer $\mathbf{q}=0$ is
\begin{align}
\Pi_{xy}(\mathbf{q}=0, \tau)&= -\frac{e^2}{L^d}\braket{ T_\tau v_x(\tau) v_y(0)} \; ,
\end{align}
where $v_\mu=(1/\hbar)\,\partial H/\partial k_\mu$ ($\mu=x$, $y$) is the velocity operator, $L$ is the linear size of systems, and $d$ is the spatial dimension. From Eq.~\eqref{eq:Green0}, the antisymmetric part of the current-current correlation function in the frequency space and at zero temperature reduce to
\begin{align}\label{eq:Pi}
&\Pi^\textrm{anti}_{xy}(\mathbf{q}=0, i\omega) \nonumber \\
&=-\frac{1}{L^d} \frac{e^2}{\hbar}\;
i\omega \, \sum_{n>0} \frac{\langle\Psi_0|v_x|\Psi_n\rangle \langle\Psi_n|v_y|\Psi_0\rangle - (x\leftrightarrow y)}{(i\omega)^2 - (E_n - E_0)^2/\hbar^2} \; .
\end{align}
Therefore, the Hall conductivity at zero temperature becomes
\begin{align}\label{eq:Hall}
\sigma_H &=\lim_{\omega\rightarrow 0} \frac{1}{\omega} \Pi^\textrm{anti}_{xy}(\mathbf{q}=0, i\omega) \nonumber \\
&=i\,\frac{e^2}{\hbar} \frac{1}{L^d}\;
\sum_{n>0} \frac{\langle\Psi_0|v_x|\Psi_n\rangle \langle\Psi_n|v_y|\Psi_0\rangle - (x\leftrightarrow y)}{(E_n - E_0)^2/\hbar^2} \; .
\end{align}

\section{Morimoto-Nagaosa model}
\label{MNmodel}

In Ref.~\cite{Morimoto-Nagaosa2016} (see also Ref.~\cite{Meng-Budich2019} for its variant), Morimoto and Nagaosa studied the influence of momentum-local two-body interactions on a typical two-band model of WSMs. Their model Hamiltonian is
\begin{equation}
H = \sum_{\mathbf{k}} \left(
\psi^\dagger_\mathbf{k} \left[ \mathbf{h}(\mathbf{k}) \cdot
\bm{\sigma} \right] \psi_\mathbf{k} + \frac{U}{2}(\psi^\dagger_\mathbf{k}\psi_\mathbf{k} - 1)^2
\right) \; ,
\end{equation}
where $\psi^\dag_{\mathbf{k}}=(c^\dag_{\mathbf{k}\uparrow},\, c^\dag_{\mathbf{k}\downarrow})$ denotes the spinor of creation operators for electrons of (pseudo-)spin $\sigma=\;\uparrow$, $\downarrow$. The first part describes a minimal two-band model of a Weyl semimetal. For simplicity, only a single pair of Weyl nodes located at momenta $\pm\mathbf{k}_\mathrm{Weyl}=(0,\,0,\,\pm k_\mathrm{Weyl})$ is assumed.
The second term represents a Hubbard-like interaction but being local in momentum space, rather than in real space.

Because different momentum states are decoupled, this interacting model can be solved exactly through diagonalization in the number representation for each $\mathbf{k}$. It is accomplished by performing a unitary transformation $\tilde{U}(\mathbf{k})$ to change to a new spinor $\phi^\dag_{\mathbf{k}}=\psi^\dag_\mathbf{k}\,\tilde{U}(\mathbf{k})
=(b^\dag_{\mathbf{k}+},\, b^\dag_{\mathbf{k}-})$ such that $\tilde{U}(\mathbf{k})^\dag [\mathbf{h}(\mathbf{k}) \cdot
\bm{\sigma}] \tilde{U}(\mathbf{k})=\sigma_z h(\mathbf{k})$
with $h(\mathbf{k})=|\mathbf{h}(\mathbf{k})|$.
For each momentum $\mathbf{k}$, the four eigenstates are $\{ \ket{0},\; b_{\mathbf{k}-}^\dag\ket{0},\; b_{\mathbf{k}+}^\dag\ket{0},\; b_{\mathbf{k}-}^\dag b_{\mathbf{k}+}^\dag\ket{0} \}$ with the corresponding energy eigenvalues $\{ U/2,\; -h(\mathbf{k}),\; h(\mathbf{k}),\; U/2 \}$, where $\ket{0}$ is the vacuum state.

For positive $U$, the many-body ground state at zero temperature and at half-filling becomes the state with a completely-filled lower band, $|\Psi_0\rangle=\prod_\mathbf{k} b_{\mathbf{k}-}^\dag\ket{0}$.
The excitation energies of the single-particle/hole excitations, $b_{\mathbf{k}+}^\dag \ket{\Psi_0}$ and $b_{\mathbf{k}-}\ket{\Psi_0}$, are nonzero, i.e., $\Delta E_{1p/1h}=h(\mathbf{k})+U/2$. The system thus behaves as a Mott insulator with a full gap.
However, we note that the bosonic charge-neutral particle-hole excitations, $b_{\mathbf{k}+}^\dag b_{\mathbf{k}-}\ket{\Psi_0}$, keep their noninteracting values of excitation energies $\Delta E_{ph}=2h(\mathbf{k})$, which reduce to zero at Weyl nodes. That is, this interacting WSM retains its \emph{gapless} Weyl nodes from the viewpoint of particle-hole excitations.
In short, from the experimental perspective, gapless Weyl nodes (or gapless excitation modes) can show up in the probes of particle-hole excitations, while they do not appear in the measurements with respect to single-particle/hole excitations (say, ARPES).

An interesting quantity, which directly relates to the particle-hole excitations, is the Hall conductivity (and thus many-body Chern number). The authors in Ref.~\cite{Morimoto-Nagaosa2016} show that its value is the same as that of a noninteracting gapless WSM. This result can be easily understood from Eq.~\eqref{kubo} or~\eqref{eq:Hall}, since the velocity operator for the Morimoto-Nagaosa model [see Eq.~\eqref{eq:v_op} but restrict to a single-orbital state], its matrix elements, and the related particle-hole excitation energies $\Delta E_{ph}$ all take the same forms as those in the noninteracting case. Therefore, the anomalous Hall effect remains unchanged as if no interaction appeared.
According to the theory of quantum Hall edge states~\cite{Wen1991,Renn1995,Yoshioka}, there must exist collective surface excitations consist of bosonic charge-neutral chiral edge modes for each two-dimensional Chern insulator of fixed $k_z$. We stress that these surface modes should be found within the particle-hole subspace and thus not of single-particle nature.

Notice that the ground state at finite $U$ is identical to that in the noninteracting case and the noninteracting correspondence is thus satisfied. Therefore, from the discussions in Ref.~\cite{Wang-Qi-Zhang2010}, the values of two topological invariants, $C(k_z)$ and $\tilde{C}(k_z)$, should be the same. This was verified by direct calculations~\cite{Morimoto-Nagaosa2016}. The nonzero values of $\tilde{C}(k_z)$ seems to be puzzling, since it is evaluated by using the single-particle Green's function (see Sec.~\ref{top_index}) which no longer contains gapless poles/excitations in the presence of $U$ as mentioned above.

The unusual result of $\tilde{C}(k_z)$ can be understood in two ways. By using the exact eigenstates and the corresponding energy spectrum, the thermal single-particle Green's function can easily be written down~\cite{Morimoto-Nagaosa2016,Meng-Budich2019}. Away from the Weyl nodes such that $h(\mathbf{k})\neq 0$, the so-called topological Hamiltonian defined in Eq.~\eqref{topo_Hami} becomes
\begin{equation}
\mathcal{H}_t = \mathbf{h}_\mathrm{eff}(\mathbf{k})\cdot\bm{\sigma} \; ,
\end{equation}
where $\mathbf{h}_\mathrm{eff}(\mathbf{k})=\mathbf{n}(\mathbf{k})[h(\mathbf{k})+U/2]$
with $\mathbf{n}(\mathbf{k})=\mathbf{h}(\mathbf{k})/h(\mathbf{k})$. Interestingly, $\mathcal{H}_t$ has the same form as the noninteracting single-particle Hamiltonian. The eigenstates $\ket{\phi_n(\mathbf{k})}$ of $\mathcal{H}_t$ are thus identical to those in the noninteracting case. Hence the topological index $\tilde{C}(k_z)$ calculated from Eq.~\eqref{topo2} becomes unchanged for nonzero $U$.
On the other hand, at the Weyl nodes with $h(\mathbf{k}=\pm\mathbf{k}_\mathrm{Weyl})=0$, we have
\begin{align}
G(\mathbf{k}=\pm\mathbf{k}_\mathrm{Weyl},i\omega) %
=\frac{i\omega}{(i\omega)^2 - (U/2)^2}\;\mathbbm{1} \; ,
\end{align}
where $\mathbbm{1}$ denotes the $2\times2$ identity matrix in the (pseudo-)spin space. We note that the zero-energy poles at $\mathbf{k}=\pm\mathbf{k}_\mathrm{Weyl}$ in the noninteracting Green's function are now replaced by the zero-energy zeros in the interacting case. Such Green's function zeros can contribute to nonzero topological index $\tilde{C}(k_z)$, as explained in Refs.~\cite{Gurarie2011,Essin-Gurarie2011}. Nevertheless, the eigenstates of the effective single-particle Hamiltonian $\mathcal{H}_t$ and the zeros in the single-particle Green's function may not correspond to real physical states even within the single-particle/hole subspace. Therefore, the topological properties inferred from the single-particle Green's function formalism can sometimes be misleading, as explained in the main text.


\end{document}